\documentclass[3p,times,twocolumn]{elsarticle}

\usepackage{ecrc}

\volume{00}

\firstpage{1}

\journalname{Nuclear Physics B Proceedings Supplement}

\runauth{Chun Shen}

\jid{nppp}

\jnltitlelogo{Nuclear Physics B Proceedings Supplement}

\usepackage{amssymb}
\usepackage[figuresright]{rotating}

\begin{document}

\begin{frontmatter}

%% Title, authors and addresses

%% use the tnoteref command within \title for footnotes;
%% use the tnotetext command for the associated footnote;
%% use the fnref command within \author or \address for footnotes;
%% use the fntext command for the associated footnote;
%% use the corref command within \author for corresponding author footnotes;
%% use the cortext command for the associated footnote;
%% use the ead command for the email address,
%% and the form \ead[url] for the home page:
%%
%% \title{Title\tnoteref{label1}}
%% \tnotetext[label1]{}
%% \author{Name\corref{cor1}\fnref{label2}}
%% \ead{email address}
%% \ead[url]{home page}
%% \fntext[label2]{}
%% \cortext[cor1]{}
%% \address{Address\fnref{label3}}
%% \fntext[label3]{}

\dochead{}
%% Use \dochead if there is an article header, e.g. \dochead{Short communication}

\title{Recent developments in the theory of electromagnetic probes in relativistic heavy-ion collisions}

%% use optional labels to link authors explicitly to addresses:
%% \author[label1,label2]{<author name>}
%% \address[label1]{<address>}
%% \address[label2]{<address>}

\author{Chun Shen}

\address{Department of Physics, McGill University, 3600 University Street, Montreal, QC, H3A 2T8, Canada}

\begin{abstract}
%% Text of abstract
The theoretical developments in the study of electromagnetic radiation in relativistic heavy-ion collisions are reviewed. The recent progress in the rates for photon and lepton pair production is discussed. Together with the improvements in the hydrodynamic descriptions of the bulk medium, the combined effort is discussed to resolve the ``direct photon flow puzzle'' in the RHIC and the LHC experiments. Further prediction of the direct photon production in high multiplicity proton-nucleus collisions at the LHC energy can serve as a signature of the quark gluon plasma formation in these small systems. Phenomenological study of dilepton production at finite net baryon density is highlighted at the collision energies available for the RHIC beam energy scan program. 
\end{abstract}

\begin{keyword}
%% keywords here, in the form: keyword \sep keyword
relativistic heavy-ion collisions, direct photons, thermal radiation, dilepton pairs
%% MSC codes here, in the form: \MSC code \sep code
%% or \MSC[2008] code \sep code (2000 is the default)

\end{keyword}

\end{frontmatter}

%%
%% Start line numbering here if you want
%%
% \linenumbers

%% main text
\section{Introduction}
\label{intro}

Electromagnetic probes, such as direct photons and dileptons ($e^+e^-$ and $\mu^+\mu^-$ pairs), are recognized as clean penetrating probes of the dense, strongly interacting medium created in relativistic heavy-ion collisions at the Relativistic Heavy-Ion Collider (RHIC) and the Large Hadron Collider (LHC). A major advantage of these electromagnetic probes over the majority of hadronic observables is that real photons and dileptons are emitted during all stages of the reaction and suffer from negligible final-state interactions. Hence, they provide almost undistorted dynamical information about their production points. 

Modeling direct photons from relativistic heavy-ion collisions requires to deal with many sources: {\it ``prompt'' photons}, produced in the earlier scattering of partons in the colliding nuclei, {\it pre-equilibrium photons}, emitted by the charge carriers as the system rapidly evolve towards local thermal equilibrium, {\it jet photons}, arising from the medium interactions and fragmentations of jets, and {\it thermal photons}, radiated from the (nearly) thermalized quark-gluon plasma (QGP) and hot hadron gas. These photons have their own characteristic momentum dependence. They can be used to diagnose the types of matter (Glasma  \cite{McLerran:2014hza}, QGP, hadron resonance gas) which make up this fireball during different stages of its evolution. 

On the other hand, recent direct photon measurements in nucleus-nucleus collisions at both RHIC and LHC energies showed large enhancements over the binary collision scaled perturbative QCD (pQCD) calculations in the low transverse momentum region \cite{Adare:2014fwh,Wilde:2012wc}. This enhanced photon production was underestimated by the thermal radiation calculated in theory. Moreover, unexpected from theory predictions, these low $p_T$ photons also carry large momentum anisotropy \cite{Adare:2011zr, Lohner:2012ct}, which is comparable to those of light hadrons. Up to today, theoretical calculations still underestimated these challenging measurements. This tension has been known as ``direct photon flow puzzle'' in our field. 

In this proceeding, I will focus on reviewing the many recent theoretical progresses that have been made to resolve this puzzle.

\section{Recent development toward resolving the direct photon flow puzzle}
\label{photon}

\subsection{Progress in thermal photon emission rates}

Our knowledge of the photon emission rates calculated from first principles have been improved along several directions over the past few years. 

The photon emission rate in a weakly-coupled, infinitely extended, static, and equilibrated QGP medium was computed to complete leading order in $\alpha_s$ almost 15 years ago \cite{Arnold:2001ms}. Recently, a full next-to-leading order (NLO) correction, $\mathcal{O}(g_s)$, was carried out in Ref. \cite{Ghiglieri:2013gia} using novel sum rules and Euclidean techniques. For the phenomenologically interesting value of $\alpha_s = 0.3$, the NLO correction represents a 20\% increase of the emission rate and has a functional form similar to the LO result. %It provides a first estimation about the associated theoretical uncertainty in calculating photon emission rate from such a medium using finite temperature perturbation theory. 

In the hot hadron gas phase, photons produced in a ($\pi$, $K$, $\rho$, $\omega$, $K^*$, $a_1$) meson gas were systematically computed in Massive Yang-Mills framework \cite{Turbide2006thesis}. Contribution from baryons was studied in \cite{Turbide:2003si} by extrapolating the medium broadened $\rho$-meson spectral function \cite{Rapp:1999ej} to zero invariant mass, This contribution includes a large set of reactions, namely radiative decays of baryon resonances, pion-baryon $t$-channel exchange, and baryon-baryon bremsstrahlung. Recently, a universal parameterization for the baryon contribution was made publicly available \cite{Heffernan:2014mla}. At relevant temperature range $T = $150-200\,MeV, the photon productions from baryon contribution can shine over those from meson gas reactions below $E_q \sim$ 1.5\,GeV \cite{Turbide:2003si, Heffernan:2014mla}. When coupled to hydrodynamic medium, the blue shift from the radial flow can enhance more photon production from baryon channels compared to meson channels \cite{vanHees:2014ida}. Baryon channels contributes about 60\% to the net hadronic thermal photon spectrum for photon energy below 3 GeV. Including these emission channels is crucial and reduces the tension with the experimental measurements. 

Because the medium created in the relativistic heavy-ion collision does not always stay in local thermal equilibrium, out-of-equilibrium medium evolution can leave traces in the final direct photon observables. Corrections to the photon production rates considering local anisotropic momentum distributions of the constituent emitters are computed for 2 $\rightarrow$ 2 scattering processes in the QGP phase \cite{Shen:2014nfa} and hadron gas phase \cite{Dion:2011pp,Shen:2014thesis}.  The relevant phenomenological studies in Ref. \cite{Dion:2011pp,Shen:2014thesis,Shen:2013cca,Shen:2014cga} showed that the out-of-equilibrium corrections from shear viscosity reduced the final direct photon anisotropy. Direct photon anisotropic flow coefficients are sensitive to the specific shear viscosity of the medium. 

\subsection{Improvement in the description of bulk medium evolution}

Thermal radiation, which tends to dominate the direct photon energy spectrum below about 2.5 GeV at RHIC and below about 3 GeV at the LHC \cite{Chatterjee:2013naa,Shen:2013vja}, is an interplay between the photon emission rates and the bulk medium evolution. The space-time volume of the fireball determines the absolute yield of thermal photon production. Hydrodynamic flow boosts the local emitted photons, changes their momentum distribution, and imprints the flow pattern on thermal photon anisotropic flow coefficients. 

The effects of initial state fluctuations coupled with event-by-event hydrodynamic evolution on direct photon observables was first investigated in Ref. \cite{Dion:2011pp,Chatterjee:2011dw}. The fluctuating hot spots in the medium enhanced the thermal photon production. Integrated event-by-event simulation framework has been developed in Ref. \cite{Shen:2014vra}.  Both direct and inclusive photon observables can be calculated within this framework \cite{Shen:2014thesis}. Through examming the extraction procedure for direct photon signal from inclusive measurements in different definitions of anisotropic flow coefficients $v_n$, scalar-product method with photon multiplicity weighting $v_n$ is preferred, which can provide an apple-to-apple comparison between theory and experiments \cite{Shen:2014lpa}.  

Recently, the state-of-the-art IP-Glasma initial conditions were coupled to a full second-order viscous hydrodynamics and UrQMD to simulate bulk dynamics of relativistic heavy-ion collisions \cite{Ryu:2015vwa}. In the hydrodynamic phase, both shear and bulk viscosities together with their second order non-linear terms are included. With this advanced model, identified particles multiplicities, mean $p_t$, and charged hadrons $v_n$ are well described from central to semi-peripheral centralities. The inclusion of bulk viscosity can increase the space-time volume by about 50\% compared the fireball evolved with only shear viscosity. The extra space-time volume increases the thermal photon production by 50\%. In the meantime, because bulk viscosity acts as a resistance to the fireball expansion, it reduces the radial flow by 10\% at the late stage of the evolution. A weaker radial flow reduces the blue shift to the thermal photons and steepens the photon spectrum. Both effects increase the thermal photon yields in the low $p_T$ regions and result in a better agreement with the experimental data \cite{Paquet:2015lta}. 

%=======================================
\begin{figure*}
\begin{tabular}{cc}
  \includegraphics[width=0.48\linewidth]{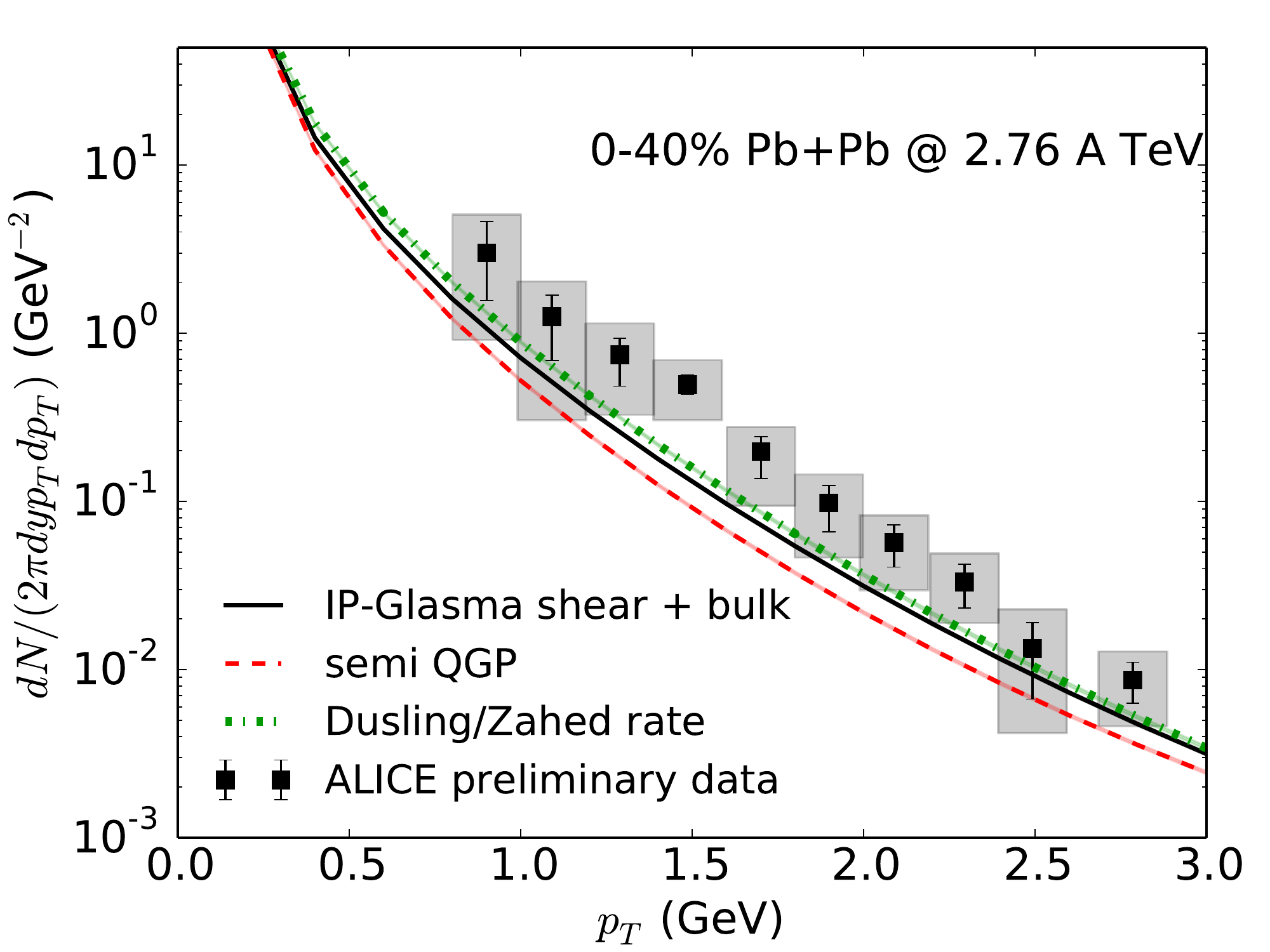} & 
  \includegraphics[width=0.48\linewidth]{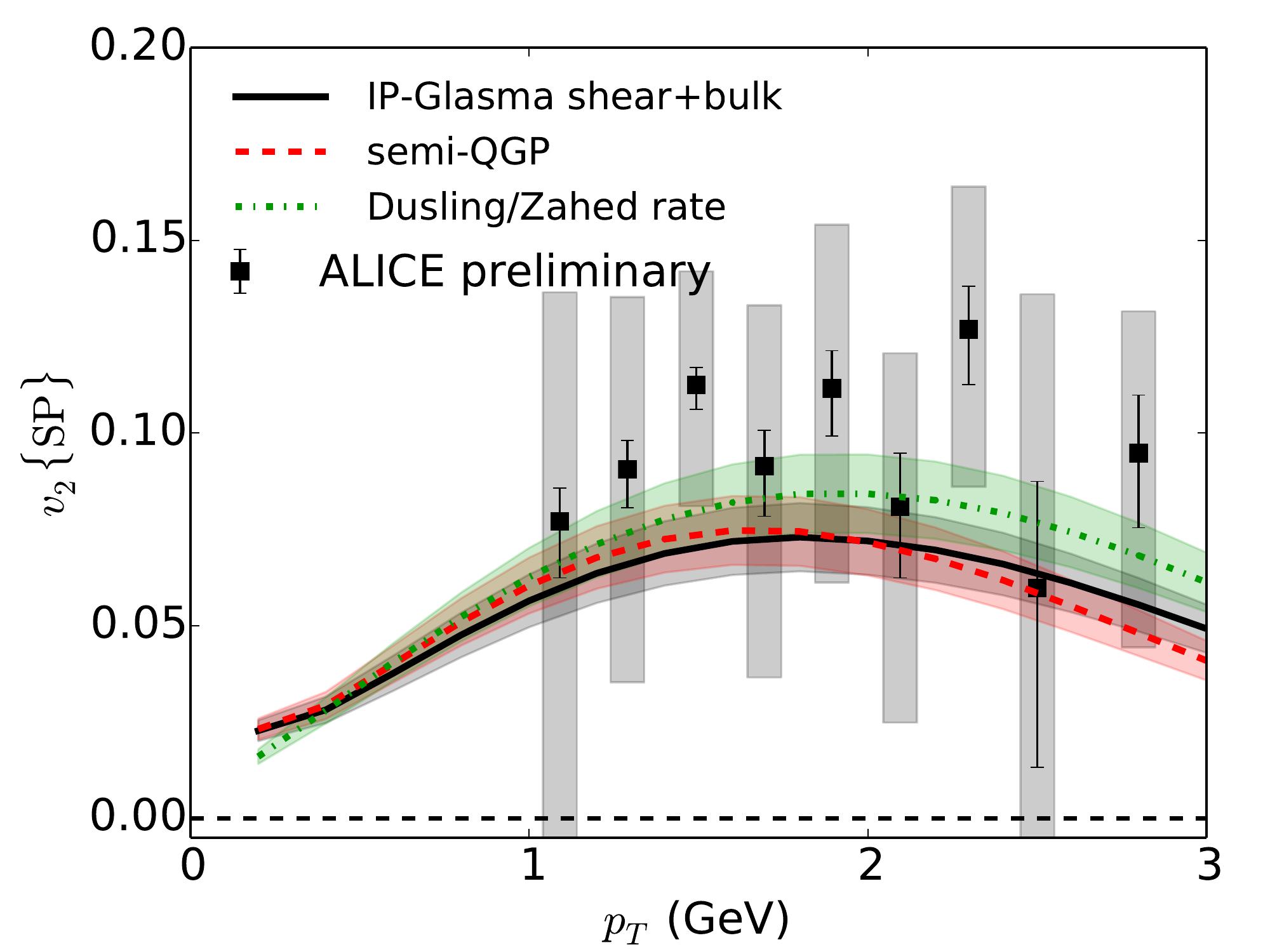}
\end{tabular}
\caption{Theoretical calculations of direct photon spectra and elliptic flow coefficient, $v_2\{\mathrm{SP}\} (p_T)$  in 0-40\% Pb+Pb collisions at $\sqrt{s_\mathrm{NN}} = 2.76$ TeV, using three different sets of photon emission rate (see text for details). \cite{Paquet:2015lta}}
\label{fig1}
\end{figure*}
%=======================================

\subsection{Photon emission near phase transition}

Photon emission rates derived in the QGP and hadron gas phases become unreliable near the phase transition region where non-perturbative physics, such as confinement, is important. In most of the phenomenology studies, one extrapolates photon emission rates calculated in both phases and matches them near $T_c$ region. Ref.~\cite{Shen:2013cca} showed that a dominant direct photon $v_2$ signal came from the phase transition region, where hydrodynamic flow has been fully developed. Thus the non-perturbative corrections to the photon emission rate are expected to be important in modeling the direct photon emission in heavy-ion collisions. The semi-QGP model uses the Polyvakov loop from Lattice QCD as an input to compute the non-perturbative corrections to the photon emission rate around the phase transition region \cite{Hidaka:2015ima,Satow:2015oha}. It results in a suppression of the QGP emission rates near $T_c$. This suppression causes a mismatch with the hadronic photon emission rates. The effects of semi-QGP suppression to the direct photon spectrum and anisotropic flow has been studied using current state-of-the-art calculation \cite{Paquet:2015lta}. Its results are shown as the red dashed lines compared with our standard calculations in Fig.~\ref{fig1}. A 30\% suppression is found in the direct photon spectra and direct photon $v_2$ has a mild increase. Hence, when confronting to explain the direct photon spectrum, the semi-QGP model faces a severer problem. 
A recent work \cite{Monnai:2015qha} on incorporating a quasiparticle picture to fit the lattice QCD EoS data shows a similar suppression of thermal photon emission near the phase transition region. %This model will face the same problem as the Semi-QGP one when compared with experimental measurements. 

In the hadron gas phase, the photon emission rate can be alternatively derived using density expansion in chiral perturbative theory \cite{Dusling:2009ej}. The leading order rate is about 40\% to 100\% larger than the rate derived in Ref. \cite{Turbide:2003si,Rapp:1999ej,Heffernan:2014mla} for $T = 100-180$ MeV. A direct channel-by-channel comparison between the two models is difficult at current stage. Using this photon emission rate for hadronic phase increases the hadronic thermal photon production, which results in a $\sim$20\% larger direct photon spectrum and $v_2(p_T)$ as shown as the green dashed curves in Fig.~\ref{fig1}.

\subsection{Comparison between transport and hydrodynamic approach}

%=======================================
\begin{figure*}
\begin{tabular}{cc}
  \includegraphics[width=0.48\linewidth]{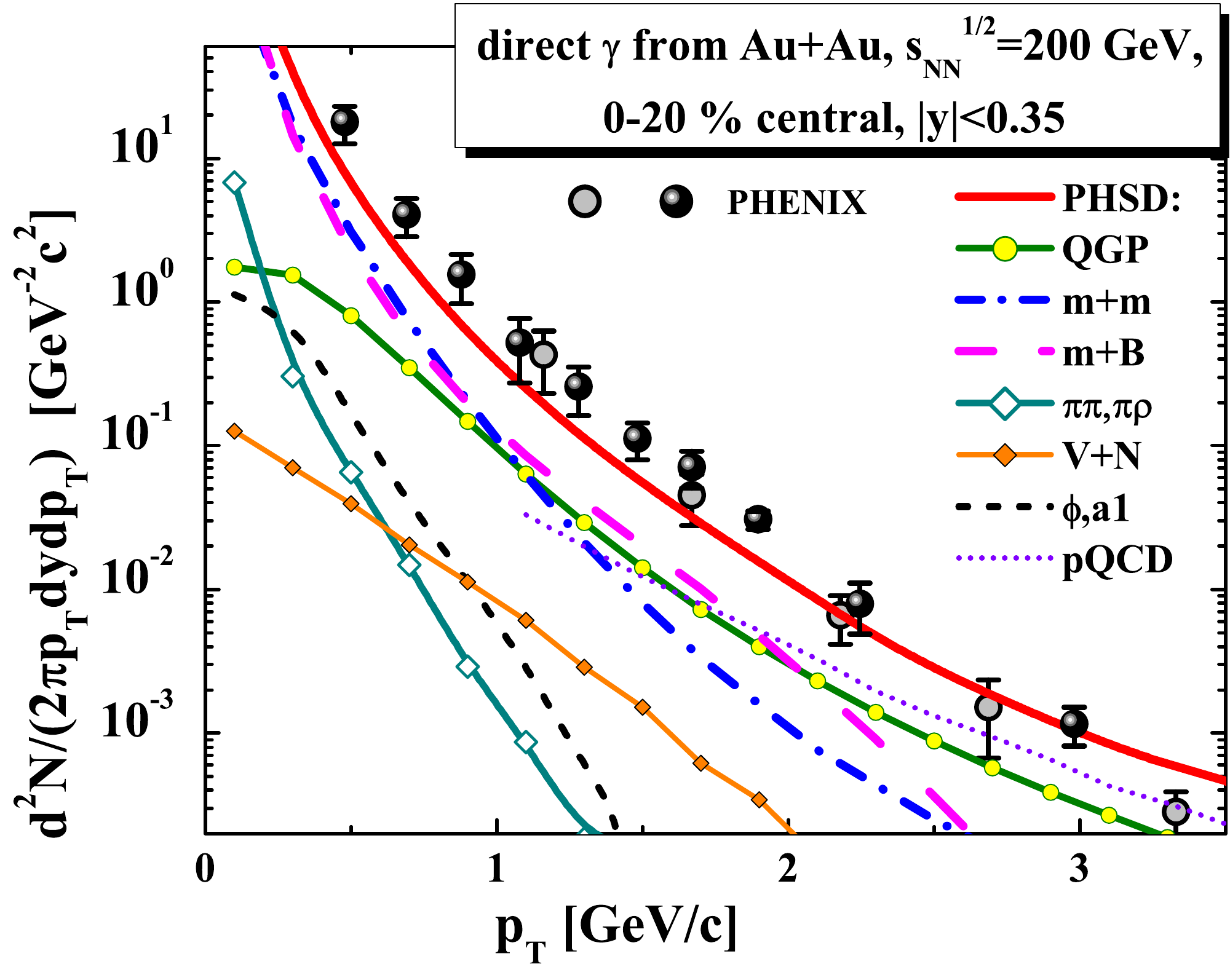} & 
  \includegraphics[width=0.48\linewidth]{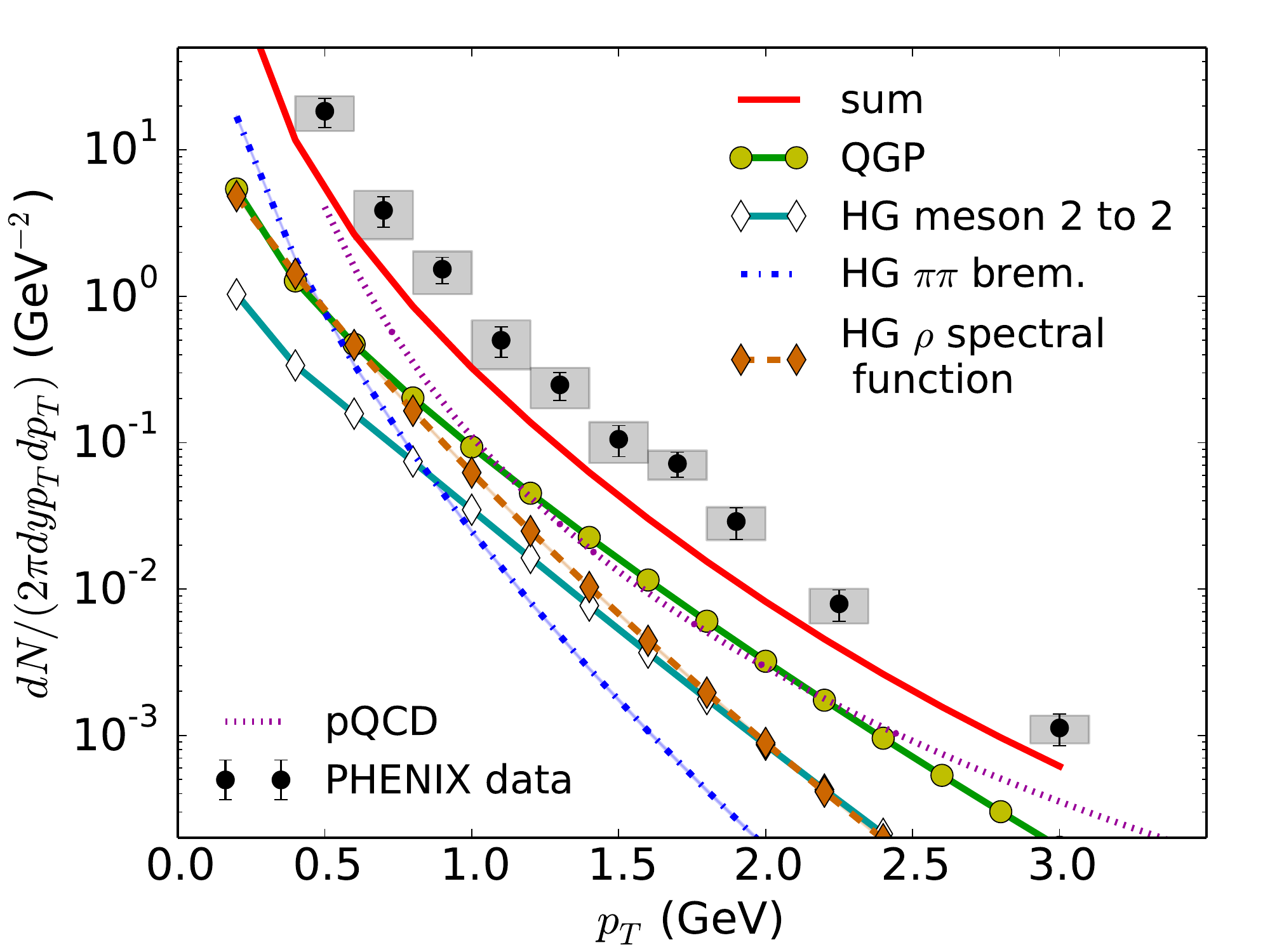}
\end{tabular}
\caption{Direct photon spectra from PHSD transport \cite{Linnyk:2015tha} and hydrodynamical simulations \cite{Paquet:2015lta} decomposed  into individual contributions in 0-20\% Au+Au collisions at $\sqrt{s_\mathrm{NN}} = 200$ GeV. Experimental data were taken from Ref.~\cite{Adare:2014fwh}.}
\label{fig2}
\end{figure*}
%=======================================

The direct photon emission is alternatively studied using a transport approach \cite{Linnyk:2015tha,Mortiz_proceeding}. A comparison of direct photon spectra between PHSD transport and hydrodynamic approaches is shown in Fig.~\ref{fig2}. The photon emissions from the QGP phase are comparable between the two approaches for $p_T > 1.0$ GeV.  However, a large difference in photon production is found in the hadronic phase. In the PHSD model \cite{Linnyk:2015tha}, most of the hadronic photons are emitted from meson-meson and meson-baryon bremsstrahlung. In the hydrodynamic simulations, photons from $\pi-\pi$ bremsstrahlung are only important for $p_T < 0.5$ GeV. A comparison of the space-time evolution may help to resolve the discrepancy between PHSD and hydrodynamic models. It can also shed light on quantifying the importance of photon emission from the dilute hadronic phase in the final measured direct photon observables. 

\subsection{Non-thermal contributions}

Besides dominant thermal photon radiation during the fireball evolution, several additional emission sources were recently proposed.

In Ref.~\cite{Campbell:2015jga}, the author proposed that photons are produced during the hadronization stage through gluon-mediated quark-anti-quark annihilations. Because a sizeable hydrodynamic flow has been generated by the hadronization stage, the photons produced from this mechanism can carry large flow anisotropies. Hence, they have the potential to increase both direct photon yields and elliptic flow. However, more theory input is needed to determine the absolute yield of photons emitted from this process. 

In Ref.~\cite{McLerran:2015mda}, the authors investigated the effect of a non-thermal (Tsallis) tail in the high $p_T$ quark distribution to photon production in the QGP phase. By coupling to a (0+1)-d Bjorken scaling medium, the authors found the averaged photon emission time can be delayed up to 1-2 fm/$c$ compared to a thermal emission source. Because the photon elliptic flow reflects the underlying flow pattern of the bulk medium at their production points, a later averaged photon emission has potential to increase the direct photon anisotropy. A calculation by embedding this effects into a more realistic (3+1)-d hydrodynamic background is needed. 

Furthermore, high energetic quarks will lose some of their energy when they pass through QGP medium. During their travelled paths, photons can be produced from both medium induced radiation and quark-photon conversion. An earlier study of the photon contribution from jet-medium interaction illustrated that the jet-medium photons were an important source in the direct photon spectra for $p_T < 4$ GeV \cite{Qin:2009bk}. Recently, a new photon production process during the jet-medium interaction was computed in Ref.~\cite{Qin:2014mya}. Implementing all these production processes and coupling them with realistic event-by-event  hydrodynamic medium is ongoing. 

Finally, because the early stage of the heavy-ion collision is a gluon dominated system, quarks and anti-quarks, who carry electric charges, are under populated. So photon production from such an out of chemically equilibrated plasma should be suppressed compared to a fully equilibrated QGP. In Ref.~\cite{Liu:2012ax, Monnai:2014kqa}, the authors found that such an early time suppression of photon emission increases the direct photon elliptic flow owing to a reduced weight of early emitted photons who carry negligible momentum anisotropy. However, as pointed out in Ref.~\cite{Gelis:2004ep}, for a gluon dominated system, its local temperature is higher than a chemically equilibrated QGP at the same energy density. This change in the EoS needs to be taken into account because it can cancel some part of the suppression from the quark fugacity factors .

\section{Thermal radiation from small collision systems}
\label{pAphoton}

%=======================================
\begin{figure}
  \includegraphics[width=1.0\linewidth]{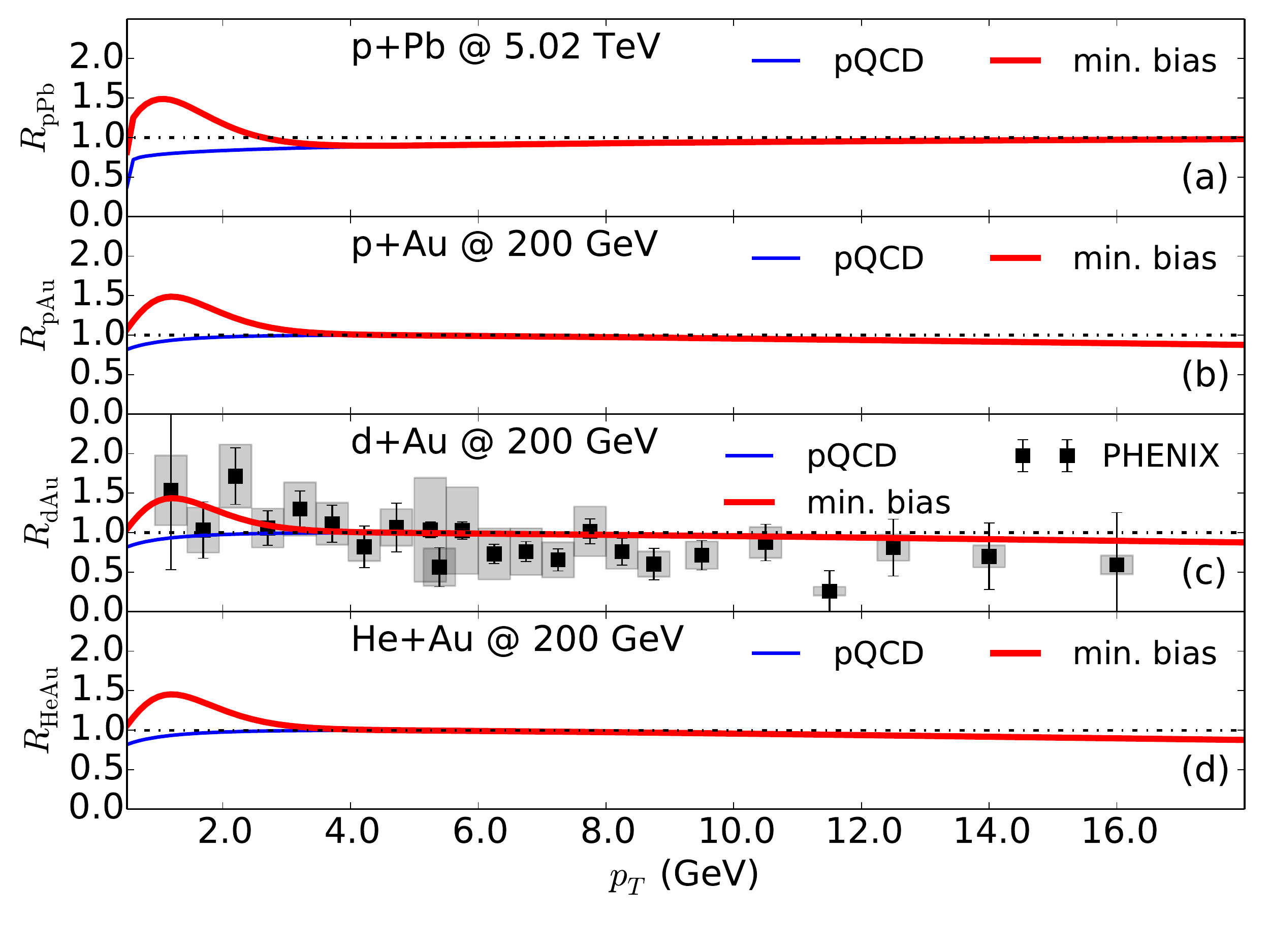} 
\caption{Direct photon $R_\mathrm{pA} (p_T)$ in minimum bias collisions in small collision systems at RHIC and LHC \cite{Shen:2015qba}. }
\label{fig3}
\end{figure}
%=======================================

While we are progressing towards the resolving the ``direct photon flow puzzle'', we also explore how to use photons as a tool to understand the nature of relativistic heavy-ion collisions. 

Unexpected collective signatures, such as multi-particle correlation and mass ordering of identified particle elliptic flow, were observed in the high multiplicity d+Au and p+Pb collisions at the RHIC and LHC. We suggested that measurements of direct photons observables in these small collision systems can provide a possible additional signature of the existence of QGP \cite{Shen:2015qba} . In 0-1\% central p+Pb collisions, thermal QGP photon production is twice compared to the prompt photons for $p_T < 2.5$\,GeV. Direct photon elliptic and triangular flow were predicted to be comparable to those in Pb+Pb collisions. Furthermore, in minimum bias collisions, thermal photons could enhance the direct photon $R_\mathrm{pA}$ by $\sim 50\%$ for $p_T < 3$ GeV as shown in Fig.~\ref{fig3}. Theory calculation is consistent with current measurement in d+Au collisions. Similar results were also obtained in p+Au and $^3$He+Au collisions at $\sqrt{s_\mathrm{NN}} = 200$\,GeV. Experimental measurements of direct photon spectra and their anisotropy flow can provide us with complementary information to determine whether the system momentum anisotropy is generated through collective expansion in these small collision systems. 

\section{Highlights of recent dilepton results}
\label{dilepton}

Virtual photons, measured as lepton pairs, are also clean and sensitive probes to all stages of the reactions. They can provide us with an additional handle at the pair's invariant mass, which allows to disentangle various sources from one to another. 

With (3+1)-d anisotropic hydrodynamic simulations, the authors of Ref. \cite{Ryblewski:2015hea, Bhattacharya:2015ada,Bhattacharya:2015wca} showed that both dileptons and real photons were sensitive probes to the non-equilibrium aspects of the early stage of heavy-ion collisions. The results were consistent with Ref. \cite{Vujanovic:2014xva, Vujanovic:2014vwa}, where hadronic emission sources were also included in the calculations. A larger initial momentum anisotropy of the system can increase dilepton yields at high invariant mass region and its elliptic flow around the $\rho$ meson mass. 

Dilepton invariant mass spectra are measured at low collision energies in the RHIC BES program \cite{Xu:2014jsa,Adamczyk:2015lme}. Because dilepton emission is sensitive to the net baryon chemical potential in the low invariant mass region, its measurement contains dynamical information of the fireball as it evolves in the QCD phase diagram. Current theoretical calculations of dilepton spectra are in agreement with the experiments \cite{Rapp:2013ema, Bratkovskaya:2011wp}. Recent study on the effect of net baryon diffusion during hydrodynamic evolution on the anisotropic flow of dileptons showed that measuring the momentum anisotropy of dileptons could further shed light on extracting the transport properties of baryon-rich QCD matter \cite{Vujanovic:2015gba, Vujanovic:2015nwv}. 

\section*{Acknowledgement}
I thank Gabriel Denicol, Charles Gale, Sangyong Jeon, Jean-Fran\c{c}ois Paquet, for collaboration and Olena Linnyk for providing the PHSD results in Fig.~\ref{fig2} and useful discussion. This work was supported by the Natural Sciences and Engineering Research
Council of Canada. 

%% The Appendices part is started with the command \appendix;
%% appendix sections are then done as normal sections
%% \appendix

%% \section{}
%% \label{}

%% References
%%
%% Following citation commands can be used in the body text:
%% Usage of \cite is as follows:
%%   \cite{key}         ==>>  [#]
%%   \cite[chap. 2]{key} ==>> [#, chap. 2]
%%

%% References with BibTeX database:
%\nocite{*}
\bibliographystyle{elsarticle-num}
\bibliography{Shen_C}

%% Authors are advised to use a BibTeX database file for their reference list.
%% The provided style file elsarticle-num.bst formats references in the required Procedia style

%% For references without a BibTeX database:

% \begin{thebibliography}{00}

%% \bibitem must have the following form:
%%   \bibitem{key}...
%%

% \bibitem{}

% \end{thebibliography}

\end{document}